# Low-temperature Benchtop-synthesis of All-inorganic Perovskite Nanowires

A. Kostopoulou,[a,*] M. Sygletou,[a] K. Brintakis,[a] A. Lappas [a] and E. Stratakis [a,b,*]

[1]Institute of Electronic Structure and Laser, Foundation for Research and Technology – Hellas, Vassilika Vouton, Heraklion, 71110, Crete, Greece
[2]Department of Materials Science and Technology, University of Crete, Vassilika Vouton, Heraklion 71003, Crete, Greece

**Abstract.** A facile, low-temperature precipitation-based method is utilized to demonstrate the synthesis of ultra-thin and highly-uniform cesium lead bromide perovskite nanowires (NWs). The reactions facilitate the NWs crystalline nature over micron-size lengths, while they impart tailored nanowire widths that range from the quantum confinement regime (~ 7 nm) and down to 2.6 nm. This colloidal synthesis approach is the first of its kind that is carried out on the work-bench, without demanding chemical synthesis equipment. Importantly, the NWs' photoluminescence is shown to become improved over time, with no tedious post-synthesis surface treatment requirement.

Inorganic semiconducting nanocrystals are unique electronic materials, exhibiting electronic structures and optoelectronic properties that are size- and shape-dependent. Their unique properties together with their easy colloidal synthesis render them efficient nanoscale functional components for multiple applications ranging from photodetectors[1] to solar energy conversion[2] and bioimaging[3].
Although a variety of synthesis procedures for semiconducting nanocrystals have been utilized, the surfactant-assisted chemical approaches hold a prominent place. These approaches have been extensively used because they can allow developing finely size-, shape-, and composition- tailored nanocrystals by careful regulation of thermodynamically/kinetically driven growth processes in liquid media.[4] So far, the most studied semiconducting nanocrystals prepared with colloidal methods have been fabricated from metal chalcogenides.[5,6] However, over the last two years, semiconducting nanocrystals of lead halide perovskite in different morphologies have shown to possess promising shape-dependent properties.[7-10] Perovskite nanocrystals have been mostly synthesized by utilizing a hot injection method in which a Schlenk line and inert gas flow are necessary. These requirements increase the material cost production, limiting their use in applications[10,11].
Recently, high-aspect-ratio perovskite nanocrystals have received considerable attention for their improved efficiency in applications such as lasers[11], photovoltaics[12] and photodetectors[13]. In 2015, the Yang group synthesized cesium lead halide NWs of >10 nm width by injecting an as- prepared Cs-oleate solution at 150 °C.[14] The reaction duration including the degassed procedure of the reactants takes more than 3 h in a fully protective atmosphere. Later, in 2016, the same group reported on thinner NWs down to 2.2 nm by a similar method at 160 °C when an additional capping agent, the l-dodecylamine was introduced.[15] The surface treatment with $PbBr_2$ precursor after the purification process is important for the quality of these NWs. At the same time, thin NWs have been synthesized by the Manna group by lowering the precursor injection temperature at 65 °C.[16] The temperature was further decreased to room temperature by Zeng group in 2017, for NWs of 1.5 nm width. [17] A special heating-treatment is required for the improvement of the PL of the NWs prepared at this temperature.

In this communication, we report on a facile precipitation-based method to synthesize thin and highly uniform $CsPbBr_3$ perovskite NWs of width 2.6 nm down to quantum confinement regime (below the Bohr radius, 7nm[7]) through a supersaturated recrystallization process at low-temperature. This methodology was carried out on the work-bench and combines the following important characteristics: i) it is a low-temperature process, ii) it is facile, i.e., there is no-need for Schleck line or inert gas flow during synthesis, iii) it is rapid and cost-efficient, iv) it is reproducible, v) it yields products of high-quality/homogeneity and vi) it utilizes commercially available starting materials. The NWs' growth and shape evolution have been carefully investigated and found that their width increases over time. Structural characterization unveiled that the NWs were highly-crystalline, adopting an orthorhombic cell from the beginning of their formation at low temperature. Optical measurements have shown a strong room-temperature photoluminescence, which gradually improves with the time that the colloidal solution is stored at room temperature, without carrying out any further purification or surface treatment.

**NWs growth and photoluminescence properties.** For the colloidal synthesis of perovskite NWs, a stock solution is first prepared with $PbBr_2$, CsBr, oleic acid and oleylamine (Oleic acid:Olam=2:1) dissolved in a certain amount of anhydrous DMF in a sealed vial closed under Ar in the protective atmosphere of a GloveBox. A small quantity of this solution is added rapidly to a sealed vial of anhydrous toluene under strong stirring, which is placed into an ice-bath. Immediately after the addition of the stock solution, a large volume of cooled anhydrous toluene was added in, for solution quench and reaction stop. The colloidal, toluene-based solution is subsequently removed from the low-temperature environment and let at room temperature without stirring. Following one day in ambient conditions, it is observed that the supernatant is consisted of compact rectangular NWs of 2.6 nm cross-section (Fig. 1 and 2b). The narrow width distribution of the NWs is indicated by the extended lamellar structures formed, assembled from individual NWs units, as well as the calculated size distribution diagrams (Figure 2b). Careful observation of an aliquot directly after the addition of the stock solution indicated a mixture of not-fully continuous nanowires and nanoplatelets (Fig. S1). Such nanoplatelets are precipitated in the reaction vial the next day.

It is observed that the NWs solution becomes photoluminescent active, following 5 minutes of storage at room temperature. To evaluate the respective optical properties, UV-Vis absorption and PL spectra were recorded and are shown in Figures 1-inset and S2. A single PL peak is observed and centered at 454 nm indicating a strong confinement effect. One of the most important consequences of the spatial confinement effect is an increase in the energy of the band-to-band excitation peaks (blue shift), as the size of the semiconducting nanoparticles is reduced in relation with the Bohr radius.[18] This emission wavelength indicates that the thickness of the NWs should be below the 7 nm Bohr exciton diameter[7]. The narrow width distribution of the NWs is indicated from the narrow PL Full-width Half Maximum (FWHM) of 22.6 nm. Analogous spectra have been recorded previously for similar well-crystalline $CsPbBr_3$ NW structures prepared with the hot injection method[15, 16] or synthesized at room temperature and treated at 120 °C (for photoluminescence activation)[17]. In particular, the PL peak shown by the Zeng group was centered at 430 nm for NWs of 1.5 nm width,[17] at 465 nm for NWs of 2.2 nm according to the Yang group[15] and at 473 nm for NWs of 3.4 nm by the Manna group.[16]

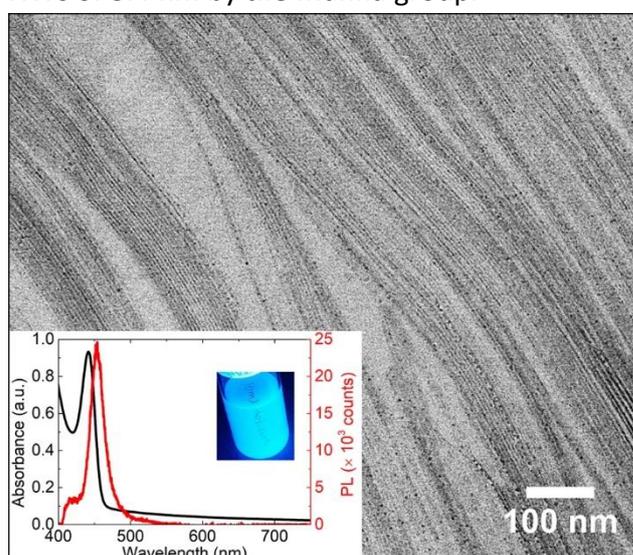

Fig. 1. Representative low-magnification TEM image of the all-inorganic lead halide NWs after one day of the toluene-based colloidal solution let at room temperature without stirring. Inset: UV-vis absorption (black line) and photoluminescence (red line) spectra; photograph of the NWs colloidal solution under UV illumination (center wavelength, λ= 365 nm).

**Morphological evolution and growth mechanism of the NWs.** To analyse the NWs formation mechanism at room temperature, different specimens from the quenched toluene-based colloidal solution have been examined in TEM. As mentioned before, when the colloidal solution is left for one day at room temperature, NWs of 2.6 nm width are formed (Figure 2b), while one week after, the NWs' width progressively increased to 6.1 nm (Fig. 2b, c). At the same time, the 1st day's NWs exhibited a narrower width distribution (Figure 2b). HRTEM analyses revealed the high-crystallinity of the synthesized wires, compatible with the orthorhombic structure of the $CsPbBr_3$ (ICSD, #97851) (Fig. 2d, e). Specifically, lattice fringes with interplanar spacings of 2.9 and 4.1 Å, which coincide with the (220) and (200) planes of the orthorhombic $CsPbBr_3$ crystal structure are measured in the HRTEM images of individual NWs. It should be noted that the NWs formed at 1 or 7 days showed the same lattice spacings and Fast
3...3

Fourier Transform (FFT) patterns (Fig. 2d, e). The crystalline nature of the NWs was also confirmed by x-ray powder diffraction (XRD) (Fig. S3). The observed Bragg reflections were nicely indexed on the basis of the orthorhombic structure of the $CsPbBr_3$ (ICSD, #97851). The distinct splitting of the diffraction peak at ~30°, which is not observed in the cubic polymorph, indicated that the NWs are predominantly adopting the orthorhombic phase. Furthermore the low-angle Bragg diffraction at 2θ= 13.1° corresponding to the (101) crystal planes is unique for the orthorhombic form and is missing in the cubic (ICSD, #29073) and in the tetragonal phases (ICSD, 109295). The orthorhombic phase is energetically stable at temperatures below 0 °C, as confirmed by earlier Raman studies, while it transforms in tetragonal at room temperature and cubic at much higher temperatures for the bulk $CsPbBr_3$.[19] Despite these, orthorhombic[14, 15, 20-22], or cubic[15, 17] crystal structures have been reported for similar NWs synthesized at room or at high temperatures (T>150 °C). Moreover, tetragonal crystal structures have been observed only for larger size morphologies, such as nanoplatelets[23] or nanosheets[24] synthesized at room temperature.

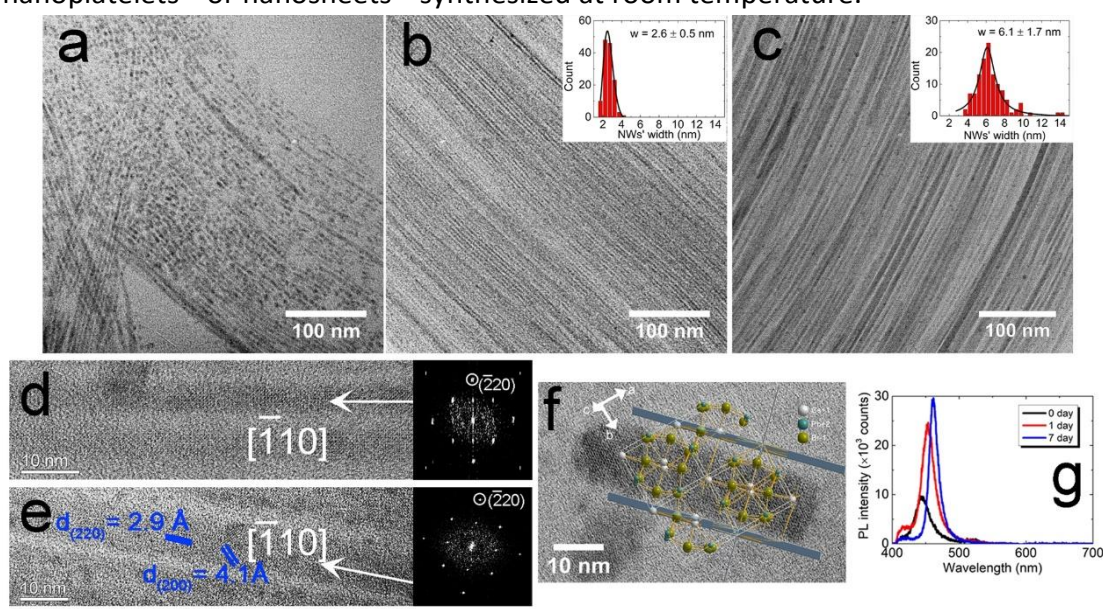

Fig. 2. Low-magnification and high-resolution TEM images of the all-inorganic lead halide NWs at 0 day (a,f), 1 day (b, d) and 7 days (c, e) after let the colloidal toluene-based solution at room temperature without stirring. The arrow indicates the growth direction of the NWs. Photoluminescence spectra (g) of the respective colloidal solutions. Insets: NWs width distributions (b, c) and the corresponding FFT patterns calculated from the HRTEM images (d, e). The zone axis in the images of panels d, e is the [001]. The indexing has been performed according to the reference pattern of the orthorhombic structure of the CsPbBr3 (ICSD, #97851). The crystal structure is visualizing with the DIAMOND software (f).[27]

Based on the FFT analysis of the relevant HRTEM images (Fig. 2d, e), the preferred growth axis of the NWs is along the [-110] direction of the orthorhombic structure. Multiple growth directions have been found in the literature for orthorhombic NWs of similar width. The [-110] found only for the NWs of 3.4 and 5.1 nm width, prepared by the hot injection method at 65 °C with Cs-oleate and a reaction duration of 200 min under Ar flow.[16] In addition, NWs of 2.2 nm width, prepared with the same method by adding the oleate-precursor at 160 °C found to be elongated along the [010] direction.[15] Recently, the [100] direction was pointed as the growth direction for NWs of 1.5 nm width synthesized at room temperature and adopting cubic structure.[17]

In order to shed light on the NWs' growth mechanism, we analysed the aliquots obtained directly after quenching of the colloidal solution with cooled toluene and the removal of the ice bath (Fig. 2a). In this case the colloids comprised small-length, bullet-like, nanorods grown along the [-110] direction of the orthorhombic structure (Fig. S4). The length of those nanorods is measured to be around 30- 40 nm, while their width is varied from ~2 to 20 nm. Such bullet-like rod morphology is more obvious in larger nanorods, as presented in Figures 2f and S6; the flatten facet on the one apex corresponds to the (110) plane. Lowering the temperature of the synthesis below room temperature slows down the growth rate, leading to the formation of perovskite particles of smaller dimensions (small nanorods, Fig. S4) compared to the particles (large nanorods and nanoplatelets) synthesized previously with similar methods.[20, 23] The growth direction of the primary-formed nanorods is the same with that of the longer NWs grown at a later time. The larger nanorods (Fig. S5) grow into platelets which are subsequently precipitated. HRTEM images, together with the respective FFT analyses (Fig. S5) revealed the high-crystallinity of the primary-formed nanorods, which is compatible with the orthorhombic structure of the CsPbBr3 (ICSD, #97851). Furthermore, weaker diffraction spots are observed which coincide with the (210) crystal planes (corresponding to lattice fringes with interplanar spacings of 3.7 Å), a unique crystallographic feature of the orthorhombic phase that is missing from the cubic and tetragonal crystal structures of the $CsPbBr_3$.

It can be postulated that, the bullet-like morphology of the primary-formed nanorods originates from strongly unidirectional growth along the [-110] direction, a process that is promoted by the intrinsic structural characteristics of the orthorhombic structure (Fig. 2f). In addition, a surface selective ligand adhesion mechanism may also operate and support this anisotropic growth. In particular, the oleic acid is anticipated to coordinate mainly with $Pb^{+2}$ ions [14] on the arrowhead apex facets of initially-formed rod-shaped seed nanocrystals, while the oleylamine could strongly bind to the $Cs^+$ ions [25] presented on the side facets. This growth mechanism is similar to that proposed by Seth and Samanta for the NWs of 70 nm in diameter and length ≥15 μm, prepared at room temperature and 40h reaction time.[20] In the later, the ratio of the ligands is much smaller (Olam: Oleic acid = 1:10) than that in our case (Olam: Oleic acid = 1:2), however playing a key role in tuning the reactivity of the monomers, and in regulating the evolution of the nanocrystal size over time in a controlled way. The surfactant molecules are continuously adsorbing and desorbing from the surface of the nanoparticles through their polar head groups, thus allowing the addition/removal of respective chemical species. The thin NWs (Fig. 2b) formed initially, grow further in thickness by attaching side to side and gradually coalesce to NWs of 6.1 nm width (Figure 2c), postulating that oriented attachment of elongated lead halide per-ovskite particles in a polar solvent[26] could be a suitable growth mechanism in the present case as well.

Upon careful inspection of the TEM micrographs of the NWs (Fig. 2 and S6), it is apparent that small nanoparticles, with a much darker contrast, are decorating their body. This is due to electron beam-induced degradation, observed during the TEM experiments when the electron beam is focused on the specimen for a long time. Such nanoparticles can be identified as Pb nanocrystals, as indicated from the respective FFT pattern obtained from an individual degraded NW (Fig. S6). Multiple weaker reflections (indicated with yellow circles in Figure S6), different from those corresponding to the orthorhombic crystal structure of the NW (indicated with red circles in Figure S6) are additionally appeared. These reflections are placed on the diffraction rings with d spacing of 2.86, 2.47, and 1.75 Å corresponding to the (111), (200) and (220) reflections of the Pb fcc cubic structure (PDF, #04-0686). Nucleation of spherical crystalline metallic Pb particles onto the NWs' surface may be the consequence of the perovskite structure degradation. Electron beam-induced degradation has been also reported before for all-inorganic lead halide nanosheets and nanocubes during TEM studies.[25]

**Self-passivation and enhanced photoluminescence behavior.** As shown in Figure 2g (red curve), the NWs exhibit a narrow PL peak, which is centered at 454 nm with a FWHM of 22.6 nm (Table S1). The NWs colloids are remarkably stable in time. Indeed, even after 7 days at room temperature, no additional PL peaks are observed at longer wavelengths, indicating the absence of NWs' aggregation. As also presented in Figure 2g (blue curve), the main PL peak at room temperature is red-shifted and becomes narrower, as the time is progressing. In particular, the PL is red-shifted by 8 nm in one week, which is in accordance with the NWs width increase. The halide-rich colloidal environment may be beneficial for obtaining NWs of better quality in such a time. Therefore, a possible self-passivation of the exposed facets, occurring at room temperature in the halide-rich colloidal environment, could explain the narrowing of the FWHM and the increase of the PL intensity. Indeed, the PL intensity enhancement observed can be attributed to improved crystallinity, giving rise to lower defect and surface electron trap densities. This colloidal behavior is stable for weeks without any surface treatment. An analogous surface passivation effect has been reported recently, to explain the enhanced quantum yields and remarkable stability of the $CsPbBr_3$ nanocubes in a halide-rich circumstances.[26]

**Parameters affecting the quality and the optical properties of the NWs**

The effects of the stirring time at low temperature (into an ice -bath) and the purification process have been investigated by combined TEM and PL analyses.

*i) Stirring time.* All the experiments have been carried out after the colloidal solution was left to rest, at room temperature, for one day (Fig. 3). Lamellar structures of well-ordered NWs are observed in the colloidal solutions prepared with a stirring time of 0 and 30 min at low temperature (Figure 3a, b); this is not the case for colloids stirred for prolonged time of 60 min (Fig. 3c). The absence of the lamellar structures in the latter case complies with the wide size distribution of the NWs. Besides this, the reduced PL intensity observed in this case as well (Fig. 3e, blue curve), complies with "peeling-off" of the attached ligands and enhanced defect density. Indeed, it has been reported that NWs without surface treatment with $PbBr_2$ are prone to breakage and ripping, during the purification process, due to ligands detachment from the NWs' surface.[15] It can be concluded that the structural and optical properties of NWs are degraded upon stirring at low temperature (Fig. 3).

***ii) Purification process.*** When the reaction solution was centrifuged at 1000 RPM for 5 min, the PL peak at 460 nm, which is characteristic for the NWs formation (Fig. S7e, black curve), is disappeared and a peak at longer wavelength (524 nm) is observed for both supernatant (Figure S7e, blue curve) and precipitated (Fig. S7e, green curve) material. This indicates the destruction of the initially formed thin NWs. Indeed, in the supernatant, broken and thicker NWs (Fig. S7a, Figure S8) as well as larger particles (Fig. S9) are observed.

Some individual particles also observed in the supernatant after the centrifugation, which confirms the degradation of the particles (Fig. S10). The precipitated particles of 100-200 nm sizes, exhibit a bulk-like and random-shaped morphology (Fig. S7b). The observed aggregation in larger particles (thick NWs and bulk-like particles) can be attributed to a respective temperature increase, due to the centrifugation process.

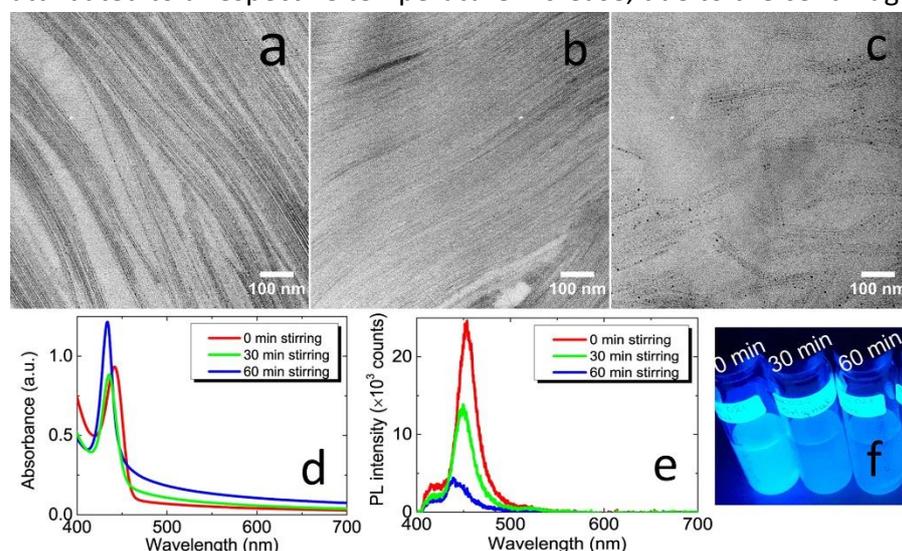

Fig. 3 Low-magnification TEM images of the NWs prepared with 0 (a), 30 (b) and 60 min (c) stirring time at low temperature. UV-vis absorption (d), and photoluminescence (e) spectra; colloidal solutions photographs (under UV illumination, with center wavelength, λ= 365 nm) (f).

In an effort to mitigate the thermal effect due to the stirring process, the vials of the original solution were placed in ice during the centrifugation process. Following this process, the NWs still appear to be ripped despite the lower temperature (Fig. S7c). However, the PL peak at 460 nm disappears, and no additional peak is observed at longer wavelengths (Fig. S7f, blue curve), in the supernatant solution, indicating limited or no formation of larger particles. While, the precipitated product is composed of large random-shaped particles (Fig. S7d). These results indicate that the NWs are present only in the supernatant of the original solution and are stable only in the toluene-based colloidal environment. It can be proposed, that the halide-rich colloidal environment allows for the protection of the NWs against degradation and promotes colloidal stability for weeks.

## Conclusions

In summary, we have reported on the colloidal synthesis of CsPbBr3 perovskite thin NWs through a rapid precipitation method at low temperature. Initially, small bullet-like nanorods have been formed at low temperature. Following one day of storage at room temperature, thin, highly uniform and micron long NWs, 2.6 nm wide have been observed, formed upon unidirectional growth along the [-110] direction of the orthorhombic crystal structure. The width of the NWs increases to 6.1 nm in one-week time through a side-to-side coalescence of the primarily formed NWs. Such NWs are measured to be highly-crystalline in orthorhombic phase, readily from the initiation of their formation at low temperature. NWs are stable in their original solution, with no need of any further surface treatment or certification process, while their PL properties progressively improve over time through a self-passivation process in a halide-rich environment.

## Conflicts of interest

There are no conflicts to declare.

## Acknowledgments

This work was supported by the EC-H2020 project, NFFA–Europe, under Grant agreement No 654360). A.K. acknowledges support from "IKY FELLOWSHIPS OF EXCELLENCE FOR POSTGRADUATE STUDIES IN GREECE-SIEMENS PROGRAM". M.S. acknowledges support from the State Scholarship Foundation (IKY) within the framework of the Action "Postdoctoral Researchers Support" (MIS: 5001552) from the resources of the OP "Human Resources Development**,** Education and Lifelong Learning"– ESPA 2014-2020 Program, contract number: 2016-050-0503-8904.

## Notes and references

# Electronic Supplementary Information

## Low-temperature Benchtop-synthesis of All-inorganic Perovskite Nanowires

A. Kostopoulou,[a,*] M. Sygletou,[a] K. Brintakis,[a] A. Lappas [a] and E. Stratakis [a,b]

[1]Institute of Electronic Structure and Laser, Foundation for Research and Technology – Hellas, Vassilika Vouton, Heraklion, 71110, Crete, Greece

[2]Department of Materials Science and Technology, University of Crete, Vassilika Vouton, Heraklion 71003, Crete, Greece

* Corresponding authors: akosto@iesl.forth.gr, stratak@iesl.forth.gr

**Chemicals:**

All reagents were of relatively high-purity. The $PbBr_2$ (trace metals basis, 99.999%), CsBr (anhydrous, 99.999%), oleic acid ($CH_3(CH_2)_7CH=CH(CH_2)_7COOH$, technical grade, 90%) and toluene (anhydrous, 99.8%) were purchased from Aldrich. The N,N-Dimethylformamide (DMF, anhydrous, 99.8%) was purchased from Alfa- Aesar and the oleylamine ($CH_3(CH_2)_7CH$-$=CH(CH_2)_7CH_2NH_2$, approximate C18-content 80-90%) from ACROS Organics. The oleic acid has been degassed for 1 hour under vacuum at 100°C and stored in the GloveBox. All the reagents were stored and handled under argon atmosphere in a glove box (MBRAUN, UNILab).

**Low-temperature synthesis of the all-inorganic perovskite nanowires:**

In a typical synthesis, a stock solution of $PbBr_2$ (0.4mmoles) and CsBr (0.4 mmoles) in 10 ml of anhydrous DMF was prepared in a sealed vial closed under Ar in protective atmosphere of a GloveBox. The solution let under stirring until the dissolution of the precursors. Then 1 ml oleic acid (0A) and 0.5 ml oleylamine (OLAm) were added in the above solution. 0.9 ml of this solution was added rapidly in 10 ml of anhydrous toluene (10ml) in a sealed vial under vigorous stirring (1000 RPM). This vial was placed in ice before the addition of the stock solution. Then, 40 ml of cooled anhydrous toluene added to the solution for quenching. All the syntheses were carried out on the bench, at 0°C without using Schlenk line and continuous Ar flow. The colloidal toluene-based solution removed from the ice and let at room temperature without stirring. A white precipitated product is observed after one day. The NWs colloidal solution is the supernatant. No further treatment or purification process is required for the characterization experiments.

**Characterization methods:**
**Structural characterization:**
*Transmission Electron Microscopy (TEM):* Low magnification and high-resolution TEM images were recorded on a LaB6 JEOL 2100 transmission electron microscope operating at an accelerating voltage of 200 kV. All the images were recorded by the GatanORIUS TM SC 1000 CCD camera. For the purposes of the TEM analysis, a drop of the as-prepared toluene-based solution was deposited onto a carbon-coated copper TEM grid and then the solution let to evaporate. Statistical analysis was carries on several HRTEM images, with the help of dedicated software (Gatan Digital Micrograph). For each sample, about 150 individual NWs were counted up. The structural features of the NWs were studied FFT patterns obtained from the HRTEM images.
*X-ray Diffraction (XRD).* Powder X-ray diffraction (XRD) studies were performed on a Rigaku D/MAX-2000H rotating anode diffractometer with Cu Kα radiation, equipped with a secondary graphite monochromator. The XRD data at room temperature were collected over a 2θ scattering range of 10–40°, with a step of 0.02° and a counting time of 10 s per step. Many drops of the as-prepared toluene-based solution was deposited onto a glassy sample holder and then the solution let to evaporate.
**Optical properties characterization:**

***Optical Absorption Spectroscopy:*** The colloidal solutions after quenching with the cooled toluene were placed in quartz cuvettes without further dilution. The UV-Vis absorption spectra were collected at room temperature on a Perkin Elmer, LAMBDA 950 UV/VIS/NIR spectrophotometer.

***Laser-Induced Fluorescence (LIF) spectroscopy:*** The colloidal solutions after quenching with the cooled toluene were placed in quartz cuvettes without further dilution. They were placed on a X-Y stage and the PL spectra were recorded at room temperature upon laser irradiation for the excitation of the samples. For sample excitation, a KrF excimer nanosecond laser, operating at 248 nm has been utilized. The pulse duration was about 20 ns, the excitation energy ~ 0.5 mJ and the laser beam diameter 3.6 mm (Fluence ~ 5 mJ/cm$^2$). The fluorescence measurements were performed at room temperature and recorded by an Andor Technology Mechelle 5000 spectrograph which is connected with an Andor iStar 734 Series, time resolved, cooled and Intensified Charge Coupled Device (ICCD). The fluorescence signal of the samples was collected and guided to the spectrograph by an optical fiber.

***Crystal structures Visualization:*** The crystal structure in the Figure S6 is prepared using *Diamond 3.0* software.[1]

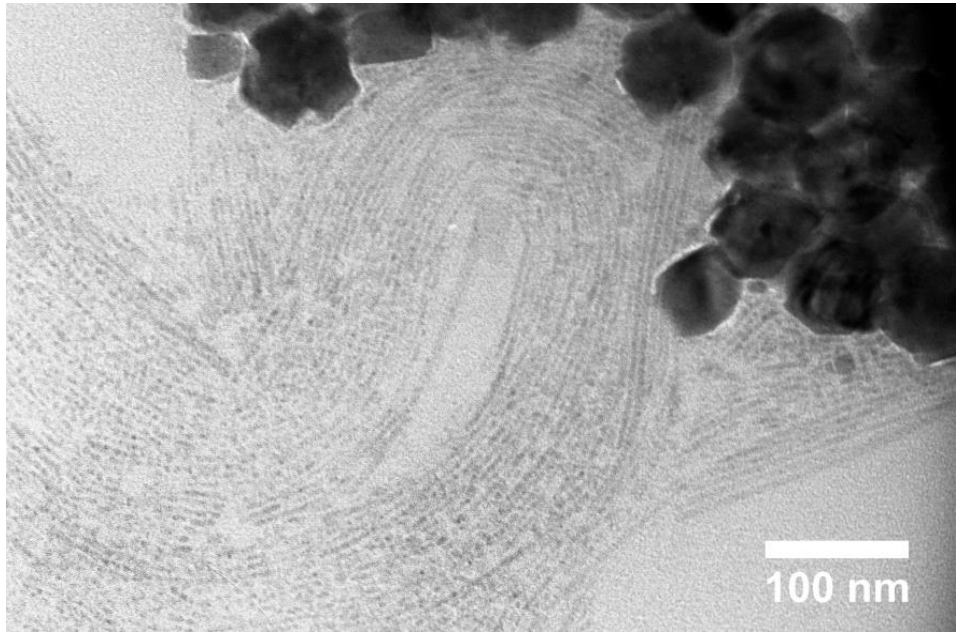

Fig. S1 Low- magnification TEM image of the mixed particle colloidal solution (discontinuous nanowires and nanoplatelets) immediately after the quenching with the cooled toluene.

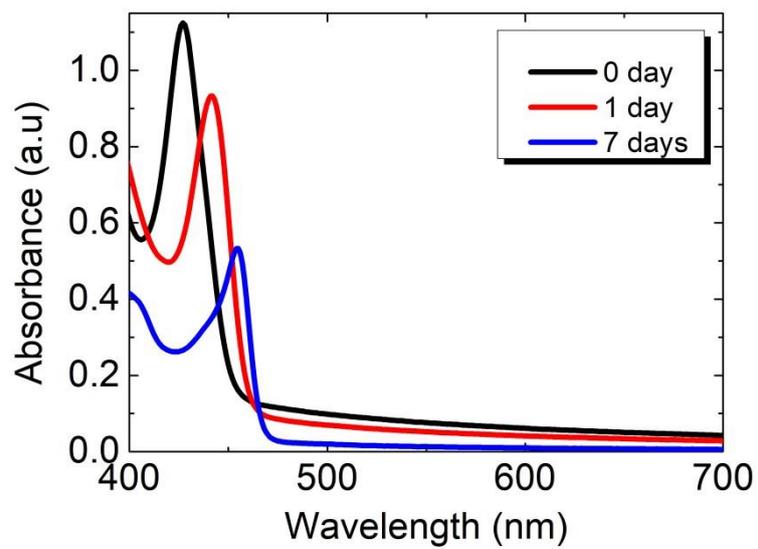

Fig. S2 UV-Vis absorption spectra of NWs' solution at different times after let the solution at room temperature without stirring.

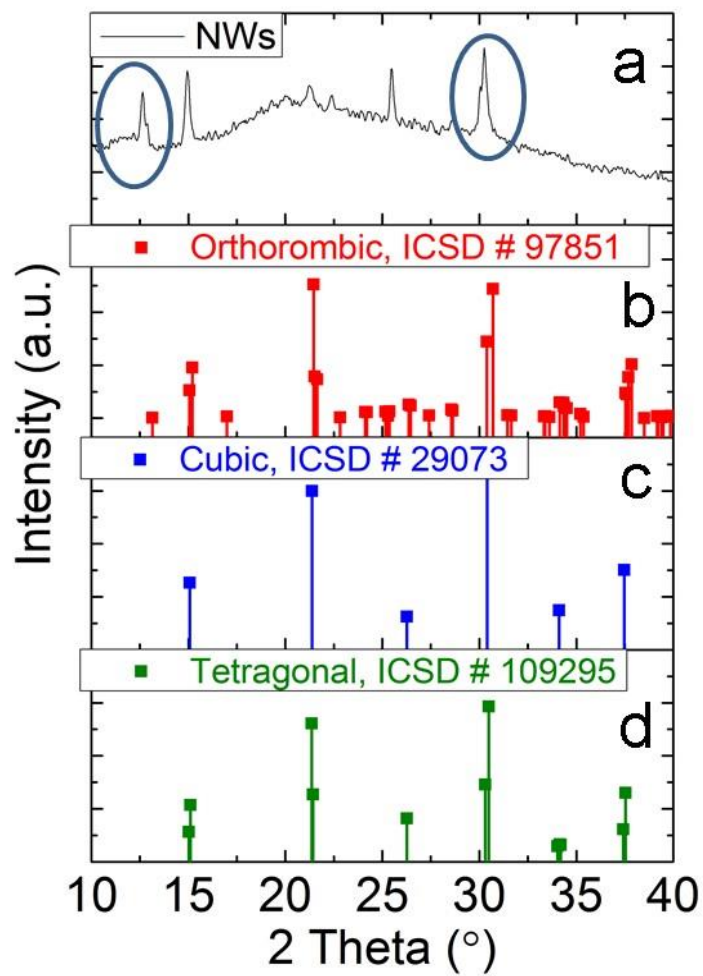

Fig. S3 Powder XRD pattern of the CsPbBr$_3$ NWs (a). Reference patterns of orthorhombic (ICSD, #97851) (b), cubic (ICSD, #29073) (c), and tetragonal (ICSD, #109295) (d) crystal structure of the CsPbBr$_3$ are provided for comparison.

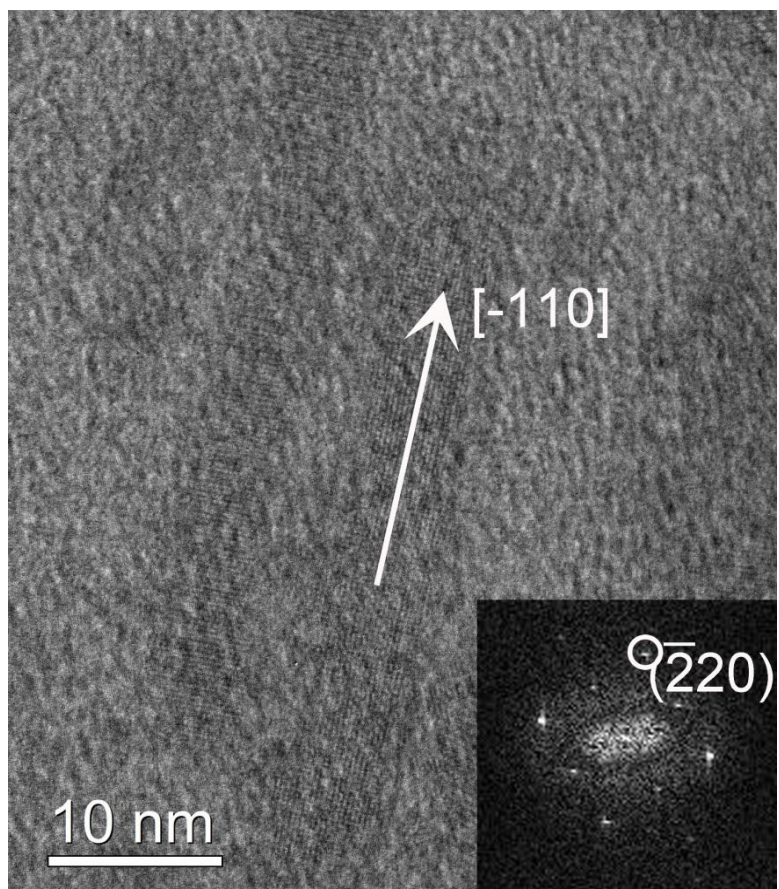

Fig. S4 HRTEM image (a) and corresponding FFT pattern (b) from small nanorods presented at the toluene-based colloidal solution at low temperature (directly after the addition of the stock precursor solution). The corresponding FFT pattern is calculated for an individual nanorod. The indexing has been performed according to the reference pattern of the orthorhombic structure of the $CsPbBr_3$ (ICSD, #97851).

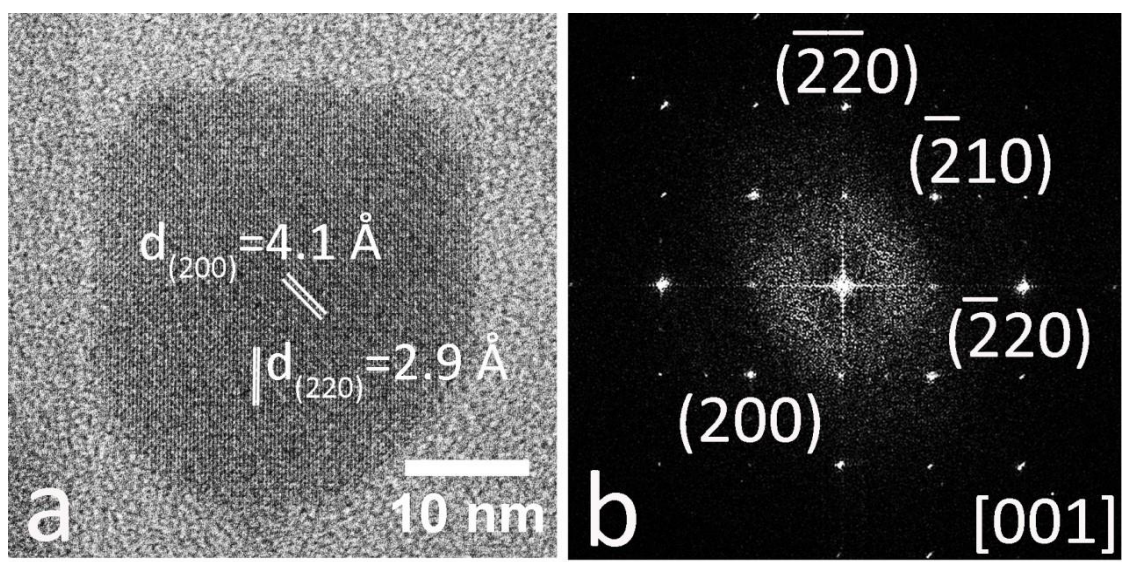

Fig. S5 HRTEM image (a) and corresponding FFT pattern (b) calculated from a thicker nanorod presented at the toluene-based colloidal solution at low temperature (directly after the addition of the stock precursor solution). The indexing has been performed according to the reference pattern of the orthorhombic structure of the $CsPbBr_3$ (ICSD, #97851).

**Electron beam-induced phase transformation-degradation process**

A thick NW, with a lot of small nanocrystals grown on it, is selected in order to have distinct reflections from the perovskite crystal structure and the possible reflections from a second material. Small Pb nanocrystals (darker contrast in the TEM images) are grown along the NWs due to the degradation of the NWs from the electron-beam irradiation through the TEM experiment.

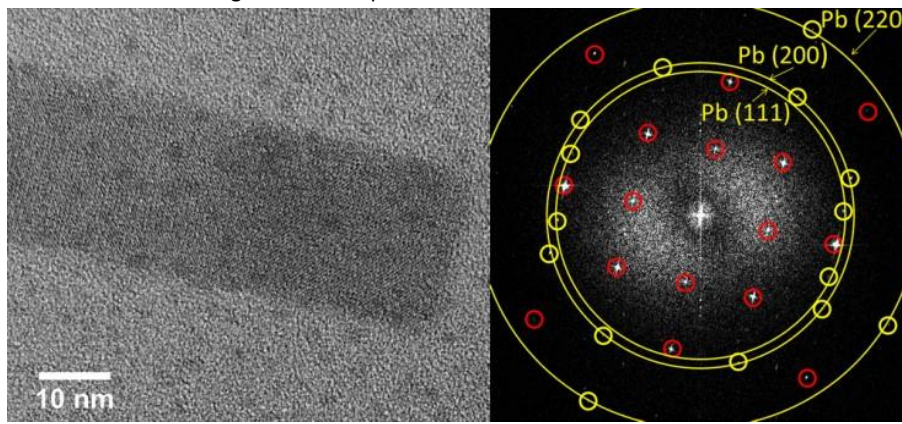

Fig. S6 HRTEM image and corresponding FFT pattern of the all-inorganic lead halide degraded NW from the electron-beam irradiation. The reflections indicated with the red circles are corresponded to the orthorhombic structure of the $CsPbBr_3$ (ICSD, #97851) while those with yellow circles to the fcc cubic structure of the metallic Pb (PDF, #04-0686).

**Table S1.** Optical parameters and NWs growth.

| Time after reaction (days) | Absorbance peak position (nm) | PL peak position (nm) | PL FWHM (nm) |
|---|---|---|---|
| 0 | 427 | 445 | 23.7 |
| 1 | 441 | 454 | 22.6 |
| 7 | 454 | 462 | 16.2 |

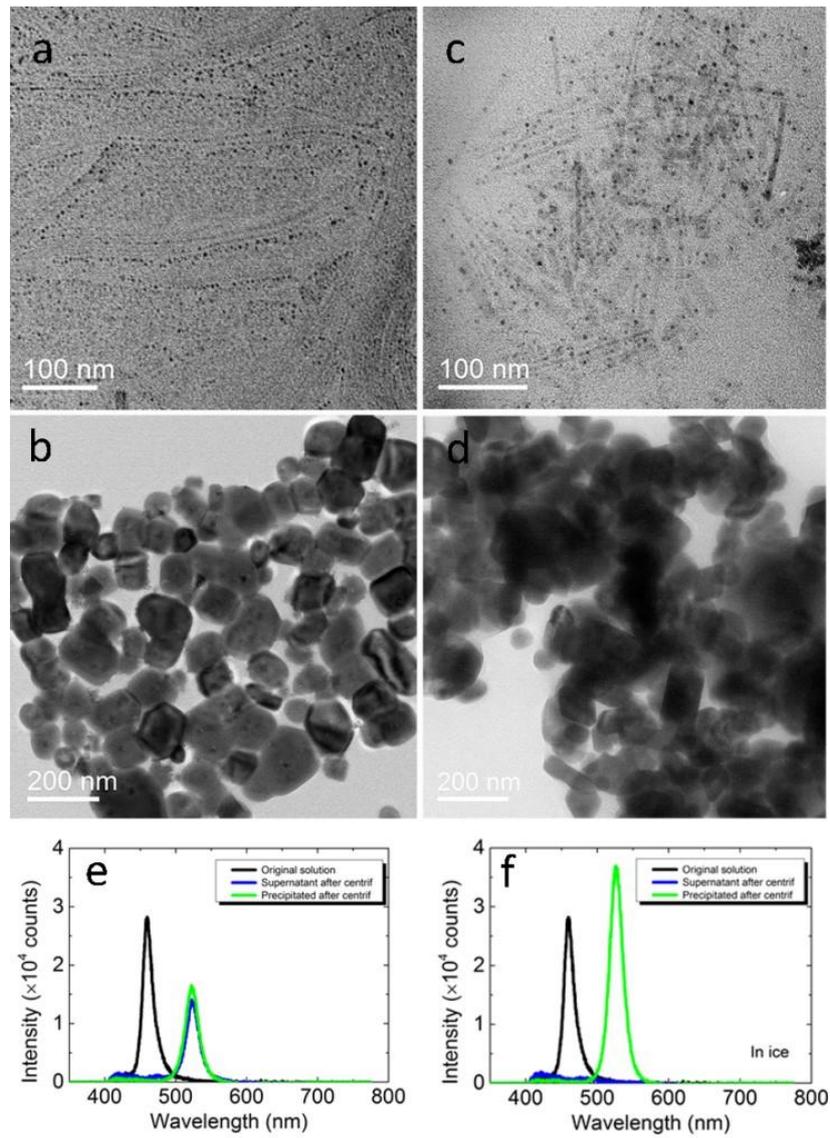

Fig. S7 Low-magnification TEM images and PL measurements of the supernatant and precipitated particles after centrifugation of the original solution (NWs and platelets) (a, b, e) and when the centrifugation done at low temperature (c, d, f).

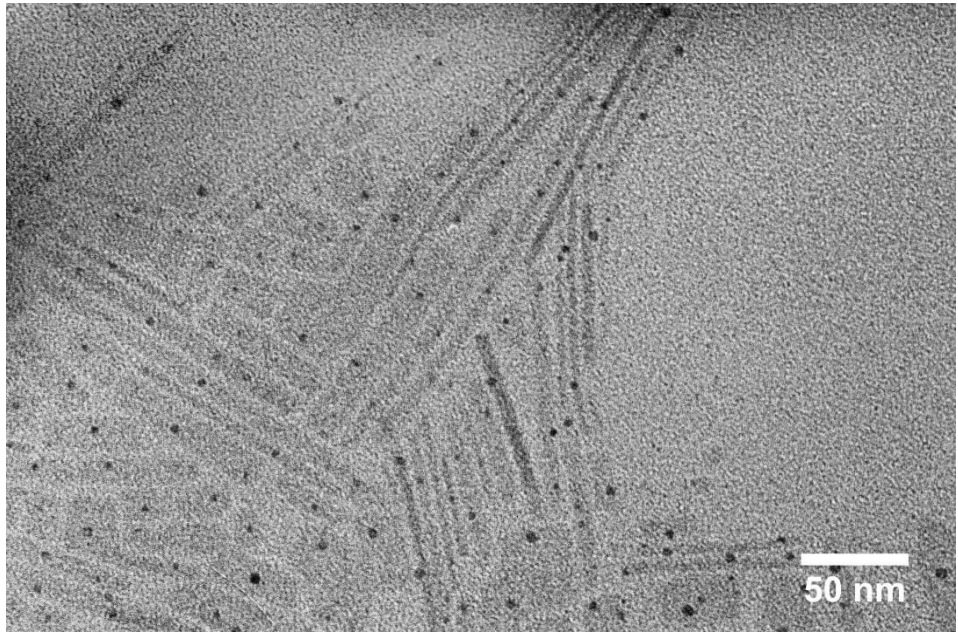

Fig. S8 Low-magnification TEM images of the NWs remained at supernatant after the centrifugation process.

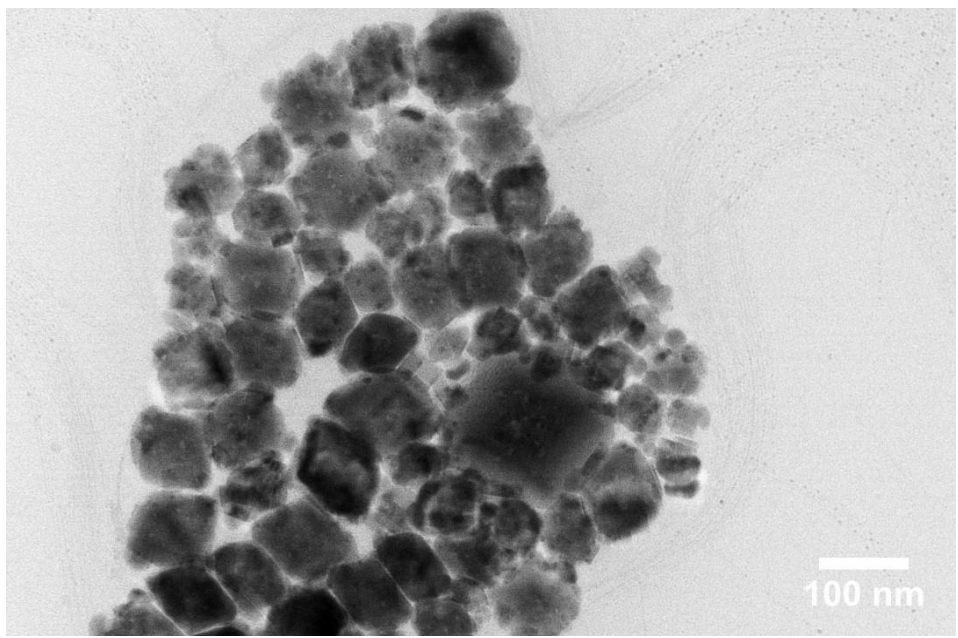

Fig. S9 Low magnification TEM images of larger particles remained in the supernatant after the centrifugation process.

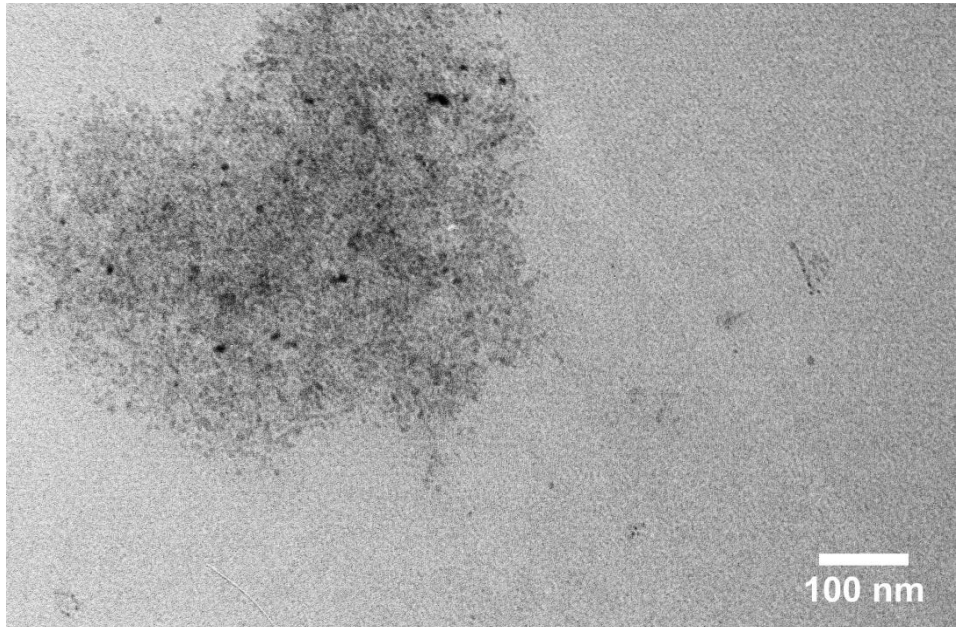

Fig. S10 Low magnification TEM images of small particles remained in the supernatant after the centrifugation process.

1.      Diamond - Crystal and Molecular Structure Visualization, Crystal Impact - Dr. H. Putz & Dr. K. Brandenburg GbR, Kreuzherrenstr. 102, 53227 Bonn, Germany, http://www.crystalimpact.com/diamond.